\documentclass[12pt]{iopart}
\usepackage{graphicx}
\newcommand\beq{\begin{equation}}
\newcommand\eeq{\end{equation}}
\newcommand\bea{\begin{eqnarray}}
\newcommand\eea{\end{eqnarray}}

\begin{document}

\title {The virial expansion of a classical interacting system}
\author{R. K. Bhaduri$^{1,2}$, M. V. N. Murthy$^2$, and Diptiman 
Sen$^3$}
\address{$^1$ Department of Physics and Astronomy, McMaster University,
Hamilton L8S 4M1, Canada}
\address{$^2$ The Institute of Mathematical Sciences, Chennai 600 113, India}
\address{$^3$ Centre for High Energy Physics, Indian Institute of Science, 
Bangalore 560 012, India}

\date{\today}

\begin{abstract}

We consider N particles interacting pair-wise by an inverse square 
potential in one dimension (Calogero-Sutherland-Moser model).  When 
trapped harmonically, its classical canonical partition function for the 
repulsive regime is known in the literature. We start by presenting a 
concise re-derivation of this result.  The equation of state is then 
calculated both for the trapped and the homogeneous gas. Finally, the 
classical limit of Wu's distribution function for fractional exclusion 
statistics is obtained and we re-derive the classical virial expansion of 
the homogeneous gas using this distribution function.

\end{abstract}

\submitto{\JPA}

\maketitle

\section{Introduction}

We consider a system of identical particles in one dimension interacting 
via an inverse square pairwise interaction. This is a class of 
integrable many-body systems known as the Calogero-Sutherland-Moser 
(CSM) model \cite{csm,poly} which is a classic example of an exactly 
solvable many-body system. For over three decades, the variants of this 
model have provided a template for analysing disparate problems in high 
energy and condensed matter physics.

The Hamiltonian of the model in the presence of a harmonic confinement is 
given by
\beq H = \sum_{i=1}^{N}[\frac{1}{2m} p_i^2 +\frac{1}{2}m\omega^2x_i^2] 
+\frac{\hbar^2\lambda}{m} \sum_{1\le i<j\le N}\frac{1}{(x_i-x_j)^2},
\label{eq1} \eeq
where $x_i, p_i$ denote the positions and momenta of the $N$ particles, 
and $\lambda$ is dimensionless coupling constant.

In the absence of a harmonic confinement, the system is classically 
integrable \cite{poly}. Defining $\hbar^2 \lambda=\alpha^2$, the integrals 
of motion are constructed using the Lax matrix defined as
\beq L_{ij} = p_i \delta_{ij} +(1-\delta_{ij})\frac{i\alpha}{x_i-x_j}.
\label{eq2} \eeq
The integrals of motion are given by
\beq I_n=tr(L^n), ~~{\rm for} ~~n=1,\ldots,N.
\label{eq3} \eeq
It is straightforward to show that the integrals of motion are in 
involution, $[I_n,I_m]=0$, and hence the system is classically integrable. 

In this paper we are interested in the classical statistical properties 
of a system of identical particles whose dynamical behaviour is described 
by the Hamiltonian in Eq. (\ref{eq1}). The quantum dynamics of such 
particles has been extensively studied \cite{csm,quantum}, and its exact 
N-particle quantum canonical partition function is known. By taking its 
$\hbar\rightarrow 0$ limit, the corresponding classical partition 
function was derived long back \cite{marchioro}. More recently, further 
studies associated with the classical integrals and related Jacobians 
have been made \cite{forrester, aomoto}. In the present paper, the 
emphasis is on the thermodynamic properties of this classical system, 
particularly in relation to the fractional exclusion statistics (FES) 
that it obeys in the quantum regime \cite{quantum,haldane}.  In sect. 2, 
we give a concise derivation of the classical N-particle canonical 
partition function, and the corresponding equation of state for the 
trapped gas.  The harmonically confined system has a constant density of 
states, and only the second virial coefficient is found to be non-zero. 
But our primary interest is in obtaining the virial expansion of the 
{\it unconfined} gas in the thermodynamic limit. The classical partition 
function for such a system is divergent. In sect. 3, we use the harmonic 
oscillator confinement as a regulator, and obtain the virial expansion 
for the equation of state in the limit of the oscillator frequency going 
to zero. In sect. 4, in the context of FES, we obtain the non-trivial 
classical distribution function by taking the appropriate limit of Wu's 
quantum occupancy factor for particles obeying fractional exclusion 
statistics \cite{haldane, wu}. Using this, we derive the virial expansion 
of the homogeneous gas and confirm the results obtained in sect. 4. We 
also find the energy of the classical CSM model at zero temperature.

\section{Classical limit of the quantum canonical partition function 
${\cal Z}_N$}

The classical partition function for $N$ identical particles is given by 
\beq Z_N(\beta) =\frac{1}{N!(2\pi\hbar)^N}\int d^Nx~d^Np ~\exp(-\beta H_N),
\label{eq4} \eeq

where the Hamiltonian $H_N$ is given by Eq. (\ref{eq1}). We choose the 
interaction strength $\hbar^2 \lambda =\alpha^2$, with $\alpha$ real. 
The interaction is therefore always repulsive. Note the explicit factor 
of $N!$ in the above expression for identical particles. While it is 
rather difficult to obtain the classical partition function by direct 
integration for all $N$, we may approach the problem as the classical 
limit of the quantum canonical partition function ${\cal Z}_N(\beta)$. 
This was the route that was taken originally in \cite{marchioro}. We 
obtain the desired result more directly, however, by using a property of 
CSM that relates to FES. To this end, we set the interaction strength 
$\lambda=g(g-1)$, where $g\geq 0$, and recall the known result \cite{JPB35}
\beq {\cal Z}_N = e^{ \hbar\beta\omega (1-g) \frac{N(N-1)}{2}} {\cal Z}_N^F,
\label{eq20} \eeq
where ${\cal Z}_N$ is for an arbitrary $g$, and ${\cal Z}_N^F$ is the 
$N$-particle non-interacting fermion partition function (for $g=1$).
Identical results are obtained when the problem is approached from the 
bosonic representation (for $g=0$).

We are now in a position to take the classical limit. We define the 
classical limit as one in which the parameters $m,\omega,\beta,\alpha$ 
are held fixed and we take the limit $\hbar \to 0$. 
Assuming this we proceed as follows. Note that we have set
\beq \alpha^2 = \hbar^2\lambda = \hbar^2 g(g-1),
\label{eq23} \eeq
where the classical interaction strength $\alpha$ is independent of $\hbar$. 
Both $\alpha$ and $g$ are positive definite and are related by 
\beq g\hbar = \frac{\hbar}{2}+\sqrt{\alpha^2+\hbar^2/4}.
\label{eq24} \eeq
Now taking the limit $\hbar \to 0$ while keeping $\alpha$ fixed implies 
$g \to \infty$ and 
\beq g\hbar \to \alpha.
\eeq
This defines the approach to the classical limit. 
Taking this limit in Eq. (\ref{eq20}) we obtain
\beq \lim_{\hbar \to 0}{\cal Z}_N = e^{-\alpha\beta\omega \frac{N(N-1)}{2}} ~
\lim_{\hbar \to 0}[{\cal Z}_N^F].
\label{eq25} \eeq

The non-interacting fermion partition function is given by the 
well-known expansion, namely,
\bea
{\cal Z}_N^F &=& (-1)^N\sum_{\cal P}\prod_l 
\frac{1}{n_l!}\left[-\frac{Z_1(l\beta)}{l}\right]^{n_l}\nonumber\\
&=&\frac{1}{N!} [{\cal Z}_1^N(\beta)-\frac{N(N-1)}{2} 
{\cal Z}_1(2\beta){\cal Z}_1^{N-2}(\beta) +\cdots],
\label{eq26} \eea
where the sum over $\cal P$ is given by the number of partitions of $N$ 
such that $\sum_{l=1}^{N} n_l l =N$ for $n_l$ and $l$ positive integers and 
\beq {\cal Z}_1 (\beta)= \frac{1}{2\sinh(\hbar\beta\omega/2)}.
\eeq
Now, taking the limit $\hbar \to 0$, we obtain the desired result
\beq \hbar^N Z_N = e^{ -\alpha\beta\omega
\frac{N(N-1)}{2}}\frac{1}{N! (\beta \omega)^N}.
\label{eq27} \eeq
This is the same result obtained in \cite{marchioro} using a 
different but longer method. 

\subsection{Equation of state of the trapped classical gas}

We can easily calculate the thermodynamic properties of the trapped
gas from $Z_N(\beta)$ given in Eq. (\ref{eq27}). The free energy is
given by $F_N=-\tau \ln Z_N$, where $\tau=1/\beta$. Since the density
of states is constant for harmonic confinement, it is like a
two-dimensional gas, and the pressure is given by 
\beq P=-\left(\frac{\partial F_N}{\partial A}\right)_{\tau},
\eeq
where $A=l^2=\hbar/m\omega$, $l$ being the oscillator length. Writing 
\beq \left(\frac{\partial F_N}{\partial A}\right)_{\tau}= \left(\frac{\partial 
F_N}{\partial\omega}\right)_{\tau} ~\left(\frac{d\omega}{d A}\right),
\eeq
and $\rho=N/A$, we obtain the equation of state 
\beq \beta P=\rho+\frac{\alpha}{2} \frac{\hbar\beta}{m} \rho^2,
\eeq
where $P$ is the pressure. Note that virial coefficients of order three 
and higher are zero.  

\section{Classical Equation of state for the homogeneous gas 
($\omega\rightarrow 0$)}
 
We now consider the virial expansion of the {\it unconfined} gas in the 
thermodynamic limit. The classical partition function for such a system 
is divergent. However, we use the harmonic oscillator confinement as a 
regulator to calculate the partition function as before and obtain the 
virial expansion for the equation of state in the limit of the 
oscillator frequency going to zero.

In the dilute limit, the equation of state is given by
\beq \beta P = \sum_{l=1}^{\infty} b_l z^l,~~ {\rm where} ~~z=e^{\beta\mu},
\label{eq46} \eeq
and $b_l$ are the cluster expansion coefficients \cite{pathria,beth} 
which appear in the fugacity expansion as above. They can 
be expressed in terms of the canonical partition functions
\beq b_l = (Z_1)^{l-1} \sum_{\{m_i\}} (-1)^{(\sum_i m_i -1)}
(\sum_i m_i -1)! \prod_i \left[\frac{Z_i}{Z_1^i}\right]^{m_i}\frac{1}{m_i !}.
\label{eq47} \eeq
Note that a cluster coefficient of order $l$ involves all the partition 
functions up to $Z_l$. The summation over $m_i$ is constrained by the 
number of partitions of $l$, that is $\sum_{i=1}^l i m_i =l.$ 

Using the expansion for the density in terms of the cluster coefficients, 
namely,
\beq \rho = \sum_{l=1}^{\infty}l b_l z^l.
\eeq
along with Eq. (\ref{eq46}), the virial expansion coefficients are defined by
\beq \beta P ~=~ \rho ~[ 1 ~+~ \sum_{n=2}^\infty a_n (\lambda_T \rho)^{n-1}],
\label{eq48} \eeq
where $\lambda_T = \sqrt{2\pi \hbar^2 \beta /m}$ is the thermal wavelength. 
The $a_k$ are the virial coefficients of the system. Note that $b_1=a_1=1$.

Consider the limit in which the confinement is removed. We
do this by taking the limit $\omega \to 0$ as follows. The cluster 
coefficients are given by, 
\bea b_2 &=& \lim_{\omega \to 0} ~\frac{Z_1}{\sqrt 2} ~[ \frac{2Z_2}{Z_1^2} ~
-~ 1], \nonumber \\
b_3 &=& \lim_{\omega \to 0} ~\frac{Z_1^2}{\sqrt 3} ~[ \frac{3Z_3}{Z_1^3} 
~-~ \frac{3Z_2}{Z_1^2} ~+~ 1].
\eea
Note that the numerical pre-factors in these expressions
for the harmonic regularisation $\omega \to 0$ are different from the 
box regularisation $L \to \infty$. In $d$ dimensions ($d=1$ in our 
case), the expressions for $b_n$ in the simple harmonic regularisation 
must be taken to be larger by a factor of $n^{d/2}$, where $n$ is the 
order of the virial coefficient, than for box regularisation. This ensures 
that they give the same result as $\omega \to 0$ and $L \to \infty$ 
respectively \cite{ouvry}.

Then the first two virial coefficients are given by
\bea a_2 &=& - ~b_2, \nonumber \\
a_3 &=& 4b_2^2 ~-~ 2 b_3 . 
\eea
For the quantum gas, we find that 
\bea b_2 &=& \frac{1}{\sqrt 2} ~(\frac{1}{2} ~-~ g), \nonumber \\
b_3 &=& \frac{\sqrt 3}{2} g(g-1) ~+~ \frac{1}{3\sqrt{3}}, 
\eea
and therefore
\bea a_2 &=& \frac{1}{\sqrt 2} (g ~-~ 1/2), \nonumber \\
a_3 &=& (2 ~-~ \sqrt{3}) ~g(g-1) ~+~ \frac{1}{2} ~-~ \frac{2}{3 \sqrt{3}}. 
\label{quvir} \eea
For $g=0$ (1), we recover the virial coefficients for a one-dimensional 
gas of non-interacting bosons (fermions). If we take the limit $\hbar \to 0$ 
and $g \to \infty$ keeping $\alpha$ fixed as before, we get the virial 
expansion for the classical gas
\beq \beta P ~=~ \rho ~[ 1 ~+~ \frac{1}{\sqrt 2} ~\sqrt{\frac{2\pi 
\beta}{m}}~ \alpha \rho ~+~ (2 ~-~ \sqrt{3}) ~\frac{2\pi \beta}{m} ~
\alpha^2 \rho^2 ~+~ \cdots]. 
\label{clvir} \eeq

\section{Classical distribution function}

An interacting system of particles described by CSM model may be mapped on to
an {\it ideal} gas obeying fractional exclusion statistics (FES) 
\cite{haldane,quantum}. In this section, we derive 
the classical virial expansion of the homogeneous gas from this starting
point and confirm that we obtain the same results as in sect. 4. It is also 
shown that the energy per unit length at zero temperature may be consistently 
obtained from our classical description by taking the limit $\hbar \to 0$.

The definition of the statistical parameter in FES, denoted by $g(>0)$, 
is based on the rate at which the number of available states in a system 
of fixed size decreases as more and more particles are added to it. The 
statistical parameter $g$ assumes the values 0 and 1 for bosons and 
fermions respectively, because the addition of one particle reduces the 
number of available states by $g$. The application of the finite 
temperature distribution function \cite{wu} then enables us to calculate 
the temperature dependent quantities of the system.

As is well known, the Haldane-Wu statistics is realized by the 
CSM model in one dimension \cite{quantum}, with the 
statistical parameter $g$ in the FES being identical to the interaction 
strength in the CSM model as noted earlier. The potential and kinetic energy 
scale in the same way in this model, and both the energy densities scale as 
$\rho^3$. The distribution function or average occupancy for FES 
particles has been derived by Wu \cite{wu} and is given by
\beq n_p = \frac{1}{w_p+g},
\label{eq28} \eeq
where $p$ denotes the momentum, and the dispersion relation is given by 
$\epsilon_p = p^2/2m$. The parameter $g$ is called the statistical 
parameter of FES since the occupancy of a given momentum state depends 
on $g$. It has been shown that the statistical parameter of FES is also 
the interaction coupling in the CSM model as used in the previous section.
The function $w_p$ satisfies the equation
\beq w_p^g(1+w_p)^{1-g}=e^{\beta(\epsilon_p-\mu)}.
\label{eq29} \eeq

We now ask the question, {\it what is the classical limit of the 
distribution function or, equivalently, how do we take the limit 
$g \to \infty$?} Let us assume that 
\beq w_p = g/\gamma_p~~ \mbox{and}~~e^{-\beta\mu}=g~e^{-\beta\mu_c},
\label{eq30} \eeq
where $\gamma_p$ depends on the momentum $p$, and $\mu_c$ is a renormalised 
chemical potential relevant to the classical limit. They also depend on 
other variables like temperature and density as will become clear below. 

Using the above, we can write Eq. (\ref{eq29}) in the following form
\beq w_p\left[\frac{w_p}{1+w_p}\right]^{g-1}=ge^{\beta(\epsilon_p-\mu_c)}.
\label{eq31} \eeq
If we now take the limit $g \to \infty$ keeping all other 
variables $\beta,\gamma_p, \epsilon_p, \mu_c$ fixed, we find that 
\beq \frac{e^{-\gamma_p}}{\gamma_p}=e^{\beta(\epsilon_p-\mu_c)},
\label{eq32} \eeq
where we have made use of the identity 
$lim_{g \to \infty}(1-\gamma_p/g)^g=e^{-\gamma_p}$. Note that 
Eq. (\ref{eq32}) uniquely fixes $\gamma_p$ in terms of the momentum $p$, 
temperature and the chemical potential since the function 
$\frac{e^{-\gamma_p}}{\gamma_p}$ monotonically goes from $\infty$ to 
0 as a function of $\gamma_p$ for $0<\gamma_p<\infty$. Furthermore, 
using Eq. (\ref{eq28}) and Eq. (\ref{eq30}) we see that
\beq n_p = \frac{1}{g}\frac{\gamma_p}{1 + \gamma_p}
\label{eq33} \eeq
which is now the analogous classical distribution function for CSM particles. 

There is yet another way of obtaining the above result from the grand 
partition function corresponding to FES, namely,
\beq \ln Z_G = \sum_p \ln (1+\frac{1}{w_p})=\sum_p \frac{\gamma_p}{g}
\label{eq34} \eeq
in the limit $g \to \infty$. We immediately find that
\beq n_p = -\frac{1}{\beta} \frac{\partial \ln Z_G}{\partial \epsilon_p}=
\frac{1}{g}\frac{\gamma_p}{1 + \gamma_p},
\label{eq35} \eeq
as desired. 

We can now obtain the classical limit by setting $g \hbar = \alpha$. In a 
system with length $L$, the free energy $F = - (1/\beta) \ln Z_G$ follows
from Eq. (\ref{eq34}):
\beq F = -\frac{L}{\beta} \int_{-\infty}^{\infty} \frac{dp}{2\pi\hbar}~ 
\frac{\gamma_p}{g} = -\frac{L}{\beta} \int_{-\infty}^{\infty} 
\frac{dp}{2\pi \alpha} ~\gamma_p. 
\label{eq36} \eeq
Since the pressure is given by $P=-(\partial F/\partial L)_{\beta}$, we see 
that
\beq \beta P = \int_{-\infty}^{\infty} \frac{dp}{2\pi \alpha} ~ \gamma_p,
\label{eq37} \eeq
The density (i.e., the number of particles per unit length) is given by
\beq \rho = \int_{-\infty}^{\infty} \frac{dp}{2\pi\hbar} ~n_p
= \int_{-\infty}^{\infty} \frac{dp}{2\pi\alpha} ~\frac{\gamma_p}{1+\gamma_p},
\label{eq38} \eeq
while the energy per unit length is
\beq E= \int_{-\infty}^{\infty} \frac{dp}{2\pi\alpha} 
\frac{\gamma_p}{1+\gamma_p} \frac{p^2}{2m}.
\label{eq39} \eeq

The virial expansion at high temperature can be obtained using Eqs. 
(\ref{eq32}), (\ref{eq37}) and (\ref{eq38}) as follows. We find that as 
$\beta \to 0$, we must take $e^{\beta \mu_c} \to 0$ so that $e^{-\beta 
(\epsilon_p - \mu_c)} \ll 1$ for all values of $p$. Using Eq. (\ref{eq32}), 
we can expand $\gamma_p$ as a power series in $e^{-\beta (\epsilon_p - 
\mu_c)}$. To go up to the third virial coefficient, we find that
\beq \gamma_p = e^{-\beta (\epsilon_p - \mu_c)} - e^{-2\beta (\epsilon_p - 
\mu_c)} + \frac{3}{2} e^{-3\beta (\epsilon_p - \mu_c)} + \cdots .
\label{eq40} \eeq
Eq. (\ref{eq38}) then gives 
\beq \rho = \frac{1}{\alpha} \sqrt{\frac{m}{2\pi \beta}} \left( e^{\beta 
\mu_c} - \sqrt{2} e^{2\beta \mu_c} + \frac{3\sqrt{3}}{2} e^{3\beta \mu_c} + 
\cdots \right).
\label{eq41} \eeq
This equation can be inverted to give
\beq e^{\beta \mu_c} = \sqrt{\frac{2\pi \beta}{m}} \alpha \rho + \sqrt{2} 
\left( \sqrt{\frac{2\pi \beta}{m}} \alpha \rho \right)^2 + (4 - 
\frac{3\sqrt{3}}{2}) \left( \sqrt{\frac{2\pi \beta}{m}} \alpha \rho 
\right)^3 + \cdots.
\label{eq42} \eeq
Eqs. (\ref{eq37}) and (\ref{eq40}) now give
\bea \beta P = \frac{1}{\alpha} \sqrt{\frac{m}{2\pi \beta}} \left( e^{\beta 
\mu_c} - \frac{1}{\sqrt{2}} e^{2\beta \mu_c} + \frac{\sqrt{3}}{2} e^{3\beta 
\mu_c} + \cdots \right).
\label{eq43} \eea
Substituting Eq. (\ref{eq42}) in (\ref{eq43}), we obtain the expression in
Eq. (\ref{clvir}).

Finally, let us consider the zero temperature limit. Note that as 
$\beta \to \infty$, $\gamma_p=0$ if $\epsilon_p > \mu_c$ and 
$\gamma_p=\infty $ if $\epsilon_p < \mu_c$; thus $\gamma_p/(1 + \gamma_p)$ 
is 0 or 1 in these two cases. This is very similar to the Fermi 
distribution function at zero temperature. Using this fact in 
Eq. (\ref{eq28}), we find that there exists a Fermi momentum $p_F = 
\sqrt{2m\mu_c}$ which is related to the density through $\rho=p_F/(\pi\alpha)$.
Eq. (\ref{eq39}) then shows that the energy per unit length is given by $E=
\pi^2\alpha^2\rho^3/(6m)$. Let us now show directly that this is the expected
value of the classical energy at zero temperature. At $T=0$, the particles
are at rest; hence the kinetic energy is zero. The repulsive two-body 
interactions in Eq. (\ref{eq1}) (where we have taken $\hbar^2 \lambda =
\alpha^2$ as usual and also set $\omega = 0$) will be minimised if the 
particles are equally spaced on a line, with the nearest neighbour spacing 
being equal to $1/\rho$. If the particles are ordered such that 
$x_i < x_{i+1}$ for all $i$, we will have $x_{i+n} -
x_i = n/\rho$. The interaction energy per particle is then given by 
\beq \frac{\alpha^2}{m} ~\sum_{n=1}^\infty ~\frac{1}{(n/\rho)^2} ~=~ 
\frac{\pi^2 \alpha^2 \rho^2}{6m}. \eeq
Thus the energy per unit length is given by $\pi^2\alpha^2\rho^3/(6m)$. 

\section{Summary}

In this paper, we have used the exact solvability of the energy spectrum of 
the quantum CSM model for any value of the interaction parameter $g$ to
study the classical limit; this limit is obtained by taking $\hbar \to 0$ 
and $g \to \infty$ keeping $g \hbar = \alpha$ fixed. 
Our derivation of $Z_N(\beta)$ is more concise than previous derivations. 
We have computed the virial expansions for
the classical CSM model with or without a harmonic confining potential
(i.e., for a homogeneous system).
Finally, we have found the classical limit of the Wu distribution function 
for FES and used this to show consistency between the virial expansions of 
the homogeneous CSM model obtained from $Z_N (\beta)$ and from FES up to
the third virial coefficient.

\ack 

We are grateful to Peter Forrester for bringing to our attention some 
earlier papers which led to a substantial revision of 
the original version of this paper. M.V.N. acknowledges the hospitality 
of Indian Institute of Science and McMaster University where parts of 
this work were done and R.K.B. acknowledges financial support by the NSERC.

\end{document}